\documentclass[12pt]{article}%
\usepackage{sw20elba}
\usepackage{amsmath}
\usepackage{graphicx}%
\usepackage{amsfonts}%
\usepackage{amssymb}

\begin{document}
\noindent{\Large HYDROSTATIC \ PRESSURE \ DEPENDENCE \ OF \ THE}%
\newline {\Large SUPERCONDUCTING \ AND \ STRUCTURAL \ PROPERTIES \ OF
\ MgB}$_{2}$\vspace{0.5cm}

\noindent J. S. Schilling,$^{a}$ J. D. Jorgensen,$^{b}$ D. G. Hinks,$^{b}$ S.
Deemyad,$^{a}$ J. Hamlin,$^{a}$ C. W. Looney,$^{c}$ and T. Tomita$^{a}$\vspace{0.2cm}

\noindent$^{a}$\textit{Department of Physics, Washington University}%
\newline \vspace{0.1cm}\noindent\textit{C.B. 1105, One Brookings Dr., St.
Louis, MO 63130}

\noindent$^{b}$\textit{Materials Science Division, Argonne National
Laboratory\newline \noindent9700 South Cass Avenue, Argonne, IL 60439}\\[0.1cm]

\noindent$^{c}$\textit{Department of Physics, Merrimack College}%
\newline \vspace{0.1cm}\noindent\textit{315 Turnpike Street, North Andover, MA
01845}\vspace{1cm}

\section{INTRODUCTION}

The past fifteen years have witnessed some of the more memorable discoveries
in the ninety-year-old field of superconductivity, including the high-$T_{c}$
oxides in 1986 \cite{n1}, the alkali-doped fullerenes in 1991 \cite{n2}, the
charge-injected fullerenes in 2000 \cite{n3}, and the binary compound
MgB$_{2}$ in January 2001 \cite{n4}. \ The discovery of superconductivity at
such high temperatures (40 K) in the simple $s,p$-metal compound MgB$_{2}$ was
quite unexpected. \ The absence \cite{n5} of the problematic weak-link
behavior of the high-$T_{c}$ oxides and the relative ease of synthesis in
various forms \cite{n6} has raised hopes that MgB$_{2}$ may be suitable for
numerous technological applications.

To aid in the search for related compounds with even better superconducting
properties and to help identify the pairing mechanism, a great deal of current
research is dedicated to fully characterizing MgB$_{2}$ in both its normal and
superconducting states. \ A wide range of experiments, including isotope
effect \cite{n7,n7'}, heat capacity \cite{n8,n8'}, inelastic neutron
scattering \cite{n9,yildirim}, NMR \cite{n9'},\ and photoemission spectroscopy
\cite{n10}, support the picture that MgB$_{2}$ is a phonon-mediated BCS
superconductor in the moderate coupling regime. \ The fact that the B isotope
effect is fifteen times that for Mg \cite{n7'} is clear evidence that the
superconducting pairing originates within the graphite-like B$_{2}$-layers,
consistent with electronic structure calculations
\cite{kortus,an,yildirim,medvedeva,kong} whereby MgB$_{2}$ is a quasi-2D
material with strong covalent bonding within the boron layers. \ The
anisotropy in the superconducting properties is appreciable, the upper
critical field ratio $H_{c2}^{ab}/H_{c2}^{c}$ reportedly being $1.7$
\cite{n12} or $2.7$ \cite{lee}, but far less than that observed in the
high-$T_{c} $ oxides \cite{n12'}. \ A full characterization of all anisotropic
properties awaits the synthesis of sufficiently large single crystals.

High pressure studies traditionally play an important role in
superconductivity. \ Even without a detailed understanding of \textit{why}
$T_{c}$ changes with pressure, a large magnitude of the pressure derivative
$dT_{c}/dP$ is a good indication that higher values of $T_{c}$ are possible at
ambient pressure through chemical means. \ It is not widely appreciated,
however, that the pressure dependence $T_{c}(P),$ like the isotope effect,
contains valuable information on the superconducting mechanism. \ In fact, in
simple $s,p$-metal BCS superconductors like Al, In, Sn, or Pb, $T_{c}$ is
found to invariably \textit{decrease} with increasing hydrostatic pressure
\cite{n13} or isotopic mass; in both cases the reduction in $T_{c}$ arises
from changes in the lattice vibration spectrum, electronic properties having
minimal effect. \ In transition-metal systems the isotope coefficient may
deviate considerably from the BCS value $\alpha=$ 0.5 and the pressure
dependence $T_{c}(P)$ is determined by changes in both lattice vibration and
electronic properties.

Soon after the discovery of superconductivity in MgB$_{2},$ three groups
reported independently that $T_{c}$ decreased under the application of high
pressure, but the rate of decrease varied considerably. \ Lorenz \textit{et
al.} \cite{n14} carried out ac susceptibility measurements in a
piston-cylinder cell to 1.8 GPa using the fluid pressure medium Fluorinert
FC77 and obtained $dT_{c}/dP\simeq$ -1.6 K/GPa. \ Saito \textit{et al.}
\cite{n14'} measured the electrical resistivity $\rho$ to 1.4 GPa using a
similar pressure technique with Fluorinert FC70 and reported the pressure
derivative -1.9 K/GPa. \ Monteverde \textit{et al.} \cite{monte} extended the
pressure range to 25 GPa in resistivity measurements in an opposed anvil cell
with solid steatite pressure medium; three of the four samples studied
exhibited widely differing pressure dependences with initial values of
$dT_{c}/dP$ ranging from -0.35 to -0.8 K/GPa. \ When pressure is applied to a
solid pressure medium like steatite, the sample is subjected to sizeable shear
stresses which may plastically deform a dense sample or compact a loosely
sintered sample, as in the experiments of\ Monteverde \textit{et al.}
\cite{monte}. \ Shear stresses are known to influence the pressure dependence
of $T_{c}$, particularly in elastically anisotropic materials such as the
high-$T_{c}$ oxides \cite{n16} or organic superconductors \cite{n17}. \ Fluid
pressure media such as Fluorinert, methanol-ethanol or silicon oil remain
fluid at RT \ to a certain pressure, but soon freeze upon cooling at
temperatures well above the superconducting transition temperature
$T_{c}\approx$ 40 K of MgB$_{2},$ thus subjecting the sample to shear
stresses, albeit relatively small ones. \ Only helium remains fluid at 40 K
for pressures to 0.5 GPa.

Since in electronic structure calculations $T_{c}$ is determined by the unit
cell dimensions and atom positions, for a quantitative comparison with theory
it is essential to complement the determination of $T_{c}(P)$ with accurate
measurements of the pressure dependence of the structure parameters. \ Effects
of pressure on the structure can affect the superconducting transition
temperature through changes in the electronic structure, phonon frequencies,
or electron-phonon coupling. \ For structures where the pressure effects are
isotropic, changes in the electronic structure are usually subtle because the
Fermi energy and features of the Fermi surface tend to simply scale together
with the cell volume. \ However, when the compression is anisotropic, as a
result of significantly different bonding strengths in different
crystallographic directions, large pressure-induced changes in the electronic
structure can occur. \ For example, in the layered copper oxide superconductor
HgBa$_{2}$CuO$_{4+x}$, where the compression is 37\% larger along the $c$ axis
than in the basal plane \cite{s1}, the pressure-induced increase in $T_{c}$
for optimally doped material is thought to occur because pressure moves a band
associated with the HgO$_{x}$ layer across the Fermi energy, creating new
carriers and ``metallizing'' the blocking layer \cite{s2,s3}. Even when such
dramatic effects do not occur, the anisotropic compression in layered
materials, such as the copper oxides, can move critical features of the
density of states (such as the van Hove singularity) with respect to the Fermi
energy resulting in changes in the carrier density\cite{s4}.

MgB$_{2}$ presents a situation where similar phenomena could occur. \ The
superconductivity is thought to result from strong electron-phonon coupling to
a particular feature of the electronic structure associated with boron
$\sigma$ bonds which lies close to the Fermi energy \cite{an}. The layered
structure of MgB$_{2}$, characterized by Mg-B bonds along the $c$ axis and B-B
bonds in the basal plane, is expected to compress anisotropically. Thus,
accurate structural data versus pressure are needed to evaluate the
pressure-induced changes in the electronic structure, as well as the changes
in phonon frequencies and electron-phonon coupling, and how these might
contribute to the pressure dependence of $T_{c}$.

Within months after the discovery of superconductivity in MgB$_{2}$, several
groups reported structural measurements versus pressure
\cite{r9,r8,r10,goncharov}. The compression is clearly anisotropic, but
quantitative agreement among the experimental results for the bulk modulus and
compression anisotropy was poor. \ Measurements made in helium gas appear to
exhibit the largest compression anistropy \{[$(dc/dP)/c_{0}$]/($da/dP)/a_{0}%
$]\} = 1.64(4) \cite{r10} and 1.9(3) \cite{goncharov}, while measurements made
in other fluids yield lower values $\sim$ 1.5 in a methanol:ethanol:water
mixture \cite{r9} and $\sim$ 1.4 in silicone oil \cite{r8}. \ Some of these
differences may be due to the degree to which the pressure fluid is truely
hydrostatic. \ Errors in the accurate \textit{in situ} measurement of lattice
parameters and the extrapolation of the results to zero pressure could also
contribute to the differences. \ Some authors \cite{monte,lorenz2} have
concluded that different samples of MgB$_{2}$ can exhibit different
pressure-dependent behavior. \ It has been speculated that samples may differ
in the amount of Mg or B vacancies, but there is no clear evidence that such
deviations in stoichiometry are possible in MgB$_{2}$. \ An alternative is
that impurity phases such as MgB$_{4}$, or elemental Mg or B, distributed at
grain boundaries or at the center of grains, modify the pressure, and amount
of shear, seen by individual crystallites of MgB$_{2}$ in sintered grains when
pressure is applied.

In this paper, we report parallel \textit{in situ }neutron powder diffraction
and $T_{c}(P)$ measurements versus pressure on the same MgB$_{2}$ sample in a
He-gas apparatus to 0.6 GPa, thus avoiding any problems with non-hydrostatic
pressure fluids or sample dependent differences. \ In addition, we present
measurements in a helium-loaded diamond-anvil-cell to 20 GPa on the same
sample. \ The high precision achieved in these measurements allows a
quantitative interpretation of the change in $T_{c}$ versus the changes in structure.

\section{EXPERIMENTAL \ METHODS}

\subsection{Sample Preparation}

The powder sample of MgB$_{2}$ for these studies was made using
isotopically-enriched $^{11}$B (Eagle Picher, 98.46 atomic \% enrichment). \ A
mixture of $^{11}$B powder (less than 200 mesh particle size) and chunks of Mg
metal was reacted for 1.5 hours in a capped BN crucible at 800$^{\circ}$C
under an argon atmosphere of 50 bar. \ As discussed below, the resulting
sample displays sharp superconducting transitions in the ac susceptibility
with full shielding. \ At ambient pressure the temperatures of the
superconducting onset and midpoint lie at 39.25 K and 39.10 K, respectively.
\ Since this sample contains isotopically pure $^{11}$B, a temperature shift
of $\Delta T_{c}\simeq0.2$ K should be added to our $T_{c}$ values before
comparing them with those from other groups using samples not isotopically
enriched ($^{10.81}$B).

\subsection{Neutron Powder Diffraction Measurements}

Neutron powder diffraction measurements were made on the Special Environment
Powder Diffractometer at the Intense Pulsed Neutron Source, Argonne National
Laboratory \cite{a1} in a He-gas pressure cell \cite{a2} at room temperature.
\ Typical data collection times were one hour at each pressure. \ Pressures
were measured continuously at the pumping station, connected to the pressure
cell by a capillary line, and are accurate and stable within 0.02 GPa. \ The
data were analyzed by the Rietveld technique using the GSAS code \cite{a3}.
\ In initial refinements, the Mg/$^{11}$B ratio was refined. \ There was no
indication of non stoichiometry within a refinement precision of about 0.5\%.
\ Fig. 1 shows the raw data and refined diffraction pattern at 0.63(2) GPa.
\ The sample is single phase and the diffraction pattern is nicely fit with
peak widths near the instrumental resolution. \ This is true at all pressures.
\ There is no evidence for any structural transitions.$T_{c}(P)$.

\subsection{Measurements in He-Gas Apparatus}

The measurements of $\ T_{c}(P)$ to 0.7 GPa were carried out using a He-gas
high-pressure system (Harwood). \ The pressure is determined by a calibrated
manganin gauge at room temperature (RT) located in the compressor system.
\ The CuBe pressure cell (Unipress) is inserted into a closed-cycle cryocooler
(Leybold) with a base temperature of 2 K and connected to the compressor
system by a 3 mm O.D. $\times$ 0.3 mm I.D. CuBe capillary tube approximately 3
m long. \ To minimize shear stresses on the sample when the helium pressure
medium freezes, a technique developed by Schirber \cite{schirber1} is applied,
whereby the top of the 15 cm long pressure cell and the capillary tube are
kept at a slightly higher temperature than the bottom so that helium freezes
from the bottom up around the sample as the pressure cell is slowly cooled (30
min) through the melting curve of helium. \ The pressure in the cell can be
changed at any temperature above the melting curve $T_{m}(P)$ of\ the helium
pressure medium (for example, $T_{m}\simeq$ 13.6 K at 0.1 GPa and $T_{m}%
\simeq$ 38.6 K at 0.50 GPa \cite{r14}). \ For pressures $P>$ 0.5 GPa, $T_{m}$
lies above the superconducting transition temperature of MgB$_{2}$ and the
sample is in frozen helium during the $T_{c} $ measurement; the slight
pressure drop (few 0.01 GPa's) on cooling in the solid helium pressure medium
from $T_{m}$ to $T_{c}$ is estimated using the known isochores of He
\cite{r14}. \ All pressures are determined at $T_{c}$.

The superconducting transition of the 8.12 mg MgB$_{2}$ powder sample is
measured by the ac susceptibility technique using a miniature
primary/secondary coil system located inside the 7 mm I.D. bore of the
pressure cell. \ An EG\&G 5210 lock-in amplifier with a transformer
preamplifier is used at 0.113 Oe (rms) field and 1,023 Hz. \ A small Pb sphere
with 1.76 mm dia (38.58 mg) is also inserted in the coil system for
susceptibility calibration purposes; for selected data the superconducting
transition temperature of this Pb sphere is used as an internal manometer
\cite{r20} to check the pressure indicated by the external manganin
gauge.\vspace{0.4cm}%

\begin{center}
\includegraphics[
natheight=11.416400in,
natwidth=18.021000in,
height=3.0424in,
width=4.7928in
]%
{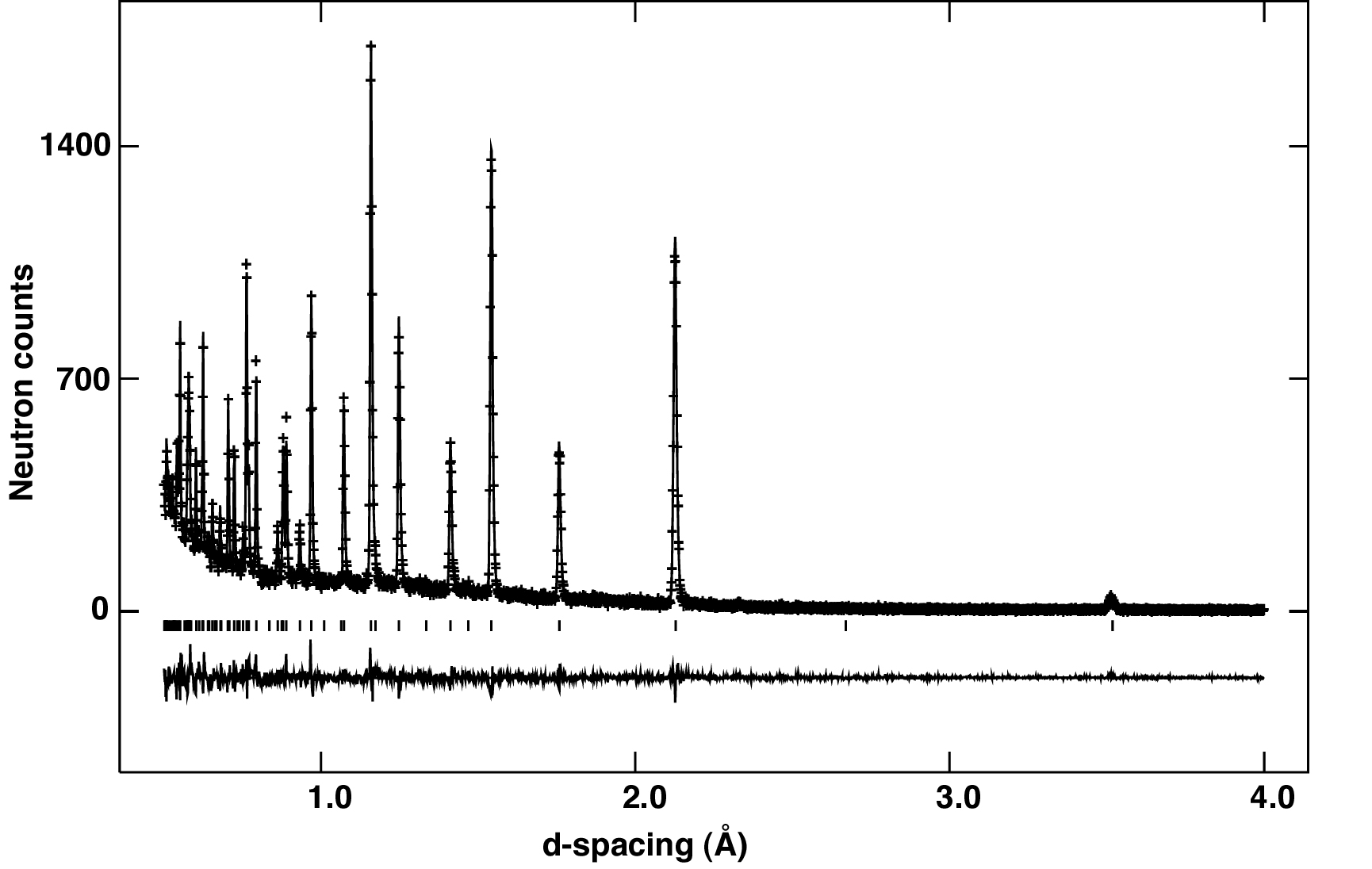}%
\end{center}

\noindent\textbf{Fig. 1.} \ Observed time-of-flight neutron powder diffraction
data and best-fit Rietveld refinement profile for MgB$_{2}$ at 0.63(2)
GPa.\ Data collection time was one hour. Crosses (+) are the raw data. The
solid line is the calculated profile.\ Tick marks indicate the positions of
all allowed reflections. A difference curve (observed minus calculated) is
plotted at the bottom.\vspace{0.4cm}

\subsection{$T_{c}(P)$ Measurements in Diamond-Anvil-Cell}

$T_{c}(P)$ can be determined to much higher pressures using a helium-loaded
diamond-anvil-cell made of hardened Cu-Be alloy fitted with 1/6-carat diamond
anvils and 0.5 mm culet diameter. \ The MgB$_{2}$ sample ($80\times80\times25$
$\mu$m$^{3}$) together with several small ruby spheres (5-10 $\mu$m dia.)
\cite{klotz} are placed in a 240 $\mu m$ dia. hole drilled through the center
of the TaW gasket. \ The pressure in the gasket hole can be changed at any
temperature from 1.6 K to RT. \ Temperature is measured by calibrated Pt and
Ge thermometers thermally anchored to the top diamond. \ The pressure in the
cell can be determined at any temperature below room temperature (RT) to
within 0.2 GPa by measuring the pressure-induced shift in the ruby R1
fluorescence line. \ The pressure is normally measured at temperatures close
to the $T_{c}$ of MgB$_{2}$.

The superconducting transition itself is determined inductively to $\pm$ 0.1 K
using two balanced primary/secondary coil systems connected to a Stanford
Research SR830 digital lock-in amplifier. \ The ac susceptibility studies were
carried out using a 3 G (r.m.s.) magnetic field at 1000 Hz. \ Over the
transition the signal changed by $\sim$ 3 nV with a background noise level of
$\sim$ 0.2 nV. \ Further details of the He-gas and diamond-anvil-cell
high-pressure techniques are given elsewhere \cite{r15,n19}.

\section{RESULTS OF EXPERIMENT}

\subsection{Pressure-Dependent Structural Properties}

The simple hexagonal structure of MgB$_{2}$ (space group P6/mmm, No. 191) is
shown in Fig. 2. The structure contains graphite-like boron layers which are
separated by hexagonal close-packed layers of metals. The center of a
hexagonal boron ring lies both directly above and below each metal.

The variation of the $a$ and $c$ lattice parameters vs. pressure is shown in
Fig. 3. Over the pressure range of this study, the changes are linear and can
be expressed as
\begin{equation}
a=a_{0}[1-0.00187(4)P]\text{ and }c=c_{0}[1-0.00307(4)P],
\end{equation}
where $a_{0}=3.08489(3)$ and $c_{0}=3.52107(5)$ are the zero-pressure lattice
parameters and $P$ is the pressure in GPa. \ Numbers in parenthesis are
standard deviations of the last significant digit. \ The bulk modulus
$[V_{0}(dP/dV)]$ obtained from these measurements is 147.2(7) GPa.

\noindent Loa and Syassen \cite{loa} used electronic structure calculations
vs. cell volume to calculate a bulk modulus of 140.1(6), in good agreement
with the experimental result. \ They also calculated the pressure dependence
of the $c/a$ ratio, getting a result in nice agreement with the observed
compression anisotropy.

The compression anisotropy, defined as $[(dc/dP)/c_{0}]/(da/dP)/a_{0}]$, is
1.64(4). \ Compression along the $c$ axis is 64\% larger than along the a
axis, consistent with the comparatively weaker (Mg-B) bonds that determine the
$c$ axis length. A similar anisotropy, but not as large, has been reported in
the refractory diboride TiB$_{2}$ \cite{s16}, which is of considerable
technological interest because of its high elastic moduli, high hardness, and
high electric conductivity. By comparison, the compression anisotropy in the
layered cuprate YBa$_{2}$Cu$_{3}$O$_{7}$ is about a factor of two \cite{a2}.
Not surprisingly, the intrinsic compression anisotropy is not observed when
pressure measurements are made in non-hydrostatic media. Recent
room-temperature x-ray diffraction measurements in diamond anvil cells using
methanol:ethanol:water \cite{r9} and silicone oil \cite{r8} as the pressure
fluids gave anisotropies of 1.5 and 1.4, respectively. \ An x-ray diffraction
study to much higher pressures using helium as the pressure fluid in a diamond
anvil cell \cite{goncharov} gives a compression anisotropy of 1.9(3), in
agreement with our result within the error bars.%

\begin{center}
\includegraphics[
natheight=6.250000in,
natwidth=6.854500in,
height=3.1306in,
width=3.4316in
]%
{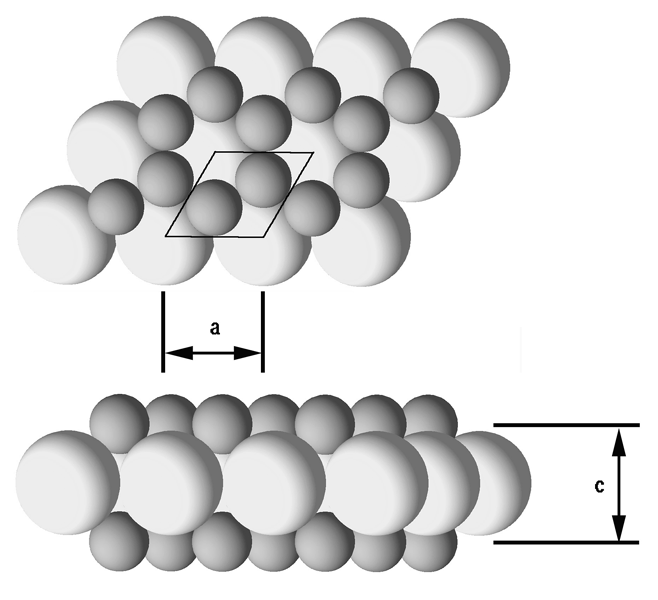}%
\end{center}

\noindent\textbf{Fig. 2.} \ Crystal structure of MgB$_{2}$ [AlB$_{2}$-type
structure; hexagonal space group P6/mmm, No. 191, with Mg at (0, 0, 0) and B
at (1/3, 2/3, 1/2)] viewed along the $c$ axis (top) and perpendicular to an
$a$ axis (bottom). \ Small spheres are B atoms; larger spheres are Mg atoms.%

\begin{center}
\includegraphics[
natheight=6.300200in,
natwidth=7.979600in,
height=3.5163in,
width=4.4486in
]%
{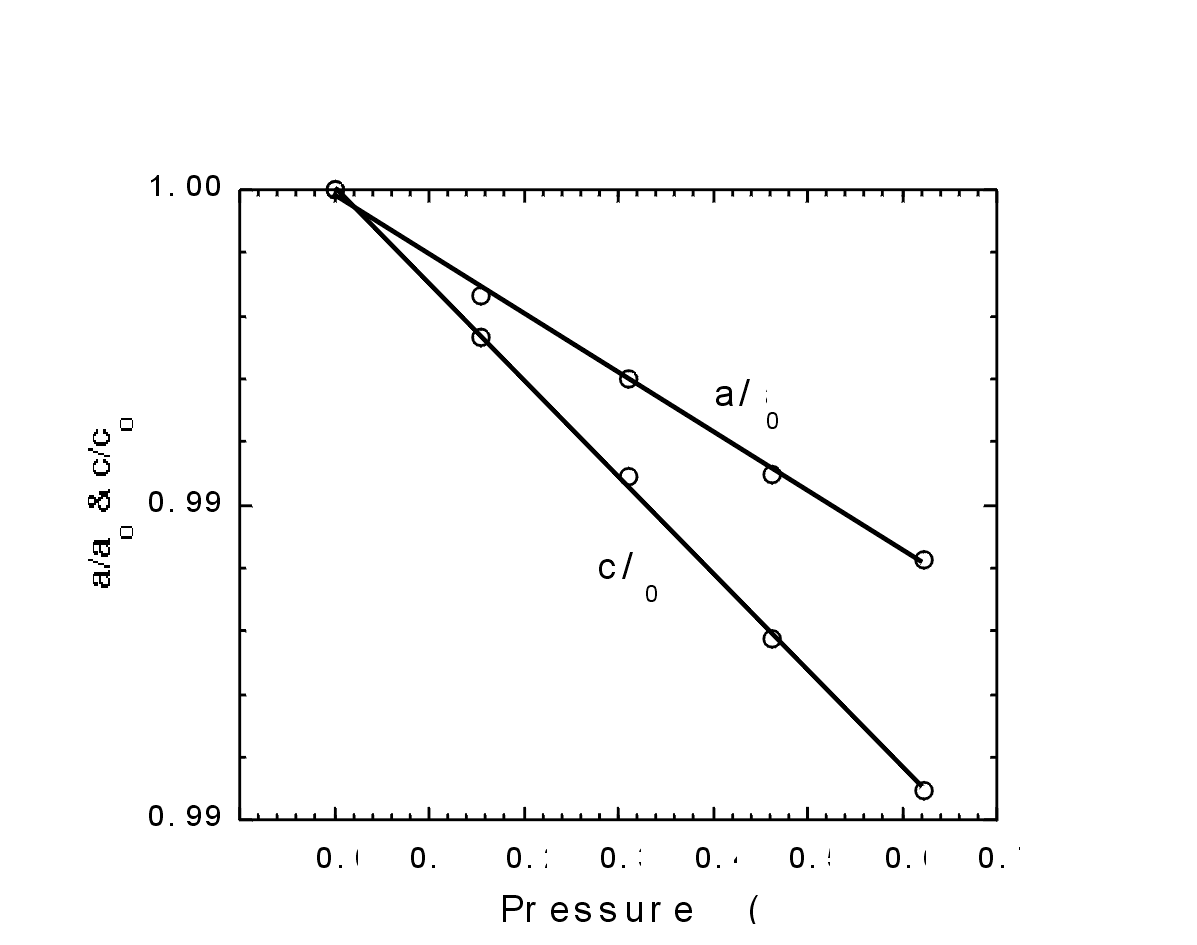}%
\end{center}

\noindent\textbf{Fig. 3.} \ Normalized $a$ and $c$ lattice parameters vs.
pressure at room temperature for MgB$_{2}$ based on neutron diffraction
measurements from Ref. \cite{r10} at five pressures using helium as the
pressure transmitting medium. \ Standard deviations of the individual points
are smaller than the symbols. \ The straight lines are linear least-sqaures
fits to the data.

\subsection{Pressure-Dependent Superconducting Properties}

\subsubsection{$T_{c}(P)$ Measurements in the He-Gas System}

In Fig. 4 we show representative examples of the superconducting transition
for MgB$_{2}$ in the ac susceptibility at both ambient and high pressure in
the He-gas system \cite{n18}. \ With increasing pressure the narrow transition
is seen to shift bodily to lower temperatures, allowing a determination of the
pressure-induced shift in $T_{c}$ to within $\pm$ 10 mK. \ Remarkably, close
inspection of the data for 0.50 GPa reveals a slight jog in the transition
curve near its midpoint, accurately marking the position of the melting curve
of helium ($T_{m}\simeq$ 38.6 K) at this pressure.%

\begin{center}
\includegraphics[
natheight=7.860300in,
natwidth=10.010200in,
height=3.288in,
width=4.1814in
]%
{figures/narlikarfigures/fig1new.png}%
\end{center}

\noindent\textbf{Fig. 4.} \ Real part of the ac susceptibility of MgB$_{2}$
versus temperature at ambient and high pressures from Ref. \cite{n18}. \ The
applied magnetic field is 0.113 Oe (rms) at 1,023 Hz. \ Intercept of straight
tangent lines defines superconducting onset at ambient pressure $T_{c}%
^{onset}(0)\simeq$ 39.25 K, with the superconducting midpoint $T_{c}%
^{mid}(0)\simeq$ 39.10 K. \ No correction is made for demagnetization
effects.\vspace{0.6cm}

In Fig. 5, the dependence of $T_{c}$ on pressure is seen to be highly linear
$dT_{c}/dP\simeq-1.11(2)$ K/GPa. \ Data were obtained following pressure
changes at both RT (unprimed data) and low temperature (primed data). \ The
dependence of $T_{c}$ on pressure thus does not depend on the
pressure/temperature history of the sample. \ Such history effects are rare in
superconductors without pressure-induced phase transitions, but do occur in
certain high-$T_{c}$ oxides containing defects with appreciable mobility at RT
\cite{relaxation}.

In selected loosely bound solids with large molecular units, like C$_{60},$
helium atoms are able to intercalate inside when pressure is applied,
diminishing the pressure-induced changes in the sample properties \cite{C60}.
\ An analysis of the MgB$_{2}$ structure readily reveals that its hexagonal
unit cell is tightly packed with insufficient space for helium atoms to
readily travel through. \ To verify that helium does not intercalate inside
MgB$_{2}$ under pressure, we carried out a parallel experiment to 0.077 GPa
with neon gas instead of helium. \ In analogy with the results on C$_{60}$
\cite{C60}$,$ the intercalation of the larger neon atoms into MgB$_{2}$ would
be more difficult than for helium. \ The fact that the pressure derivative
$dT_{c}/dP$ is the \textit{same} for both helium and neon confirms the absence
of intercalation effects in the present experiments.%

\begin{center}
\includegraphics[
natheight=7.780700in,
natwidth=9.476600in,
height=3.5137in,
width=4.2739in
]%
{figures/narlikarfigures/fig2new.png}%
\end{center}

\noindent\textbf{Fig. 5.} \ Superconducting transition temperature onset
versus applied pressure from Ref. \cite{n18}. \ Numbers give order of
measurement. \ Data for pts. 2$^{\prime}$, 6, 8$^{\prime}$, and 11 are shown
in Fig. 4. \ A typical error bar for $T_{c}$ ($\pm0.01$ K) is given in lower
left corner; the error in pressure is less than the symbol size. \ Pressure
was either changed at RT (unprimed numbers) or at low temperatures $\sim$ 60 K
(primed numbers).\vspace{0.6cm}

Since for pressures less than 0.5 GPa the sample is surrounded by fluid helium
during the $T_{c}$ measurement, the measured slope $dT_{c}/dP\simeq$ -1.11
K/GPa to this pressure gives the true hydrostatic pressure dependence for
MgB$_{2}$. \ For $P>0.5$ GPa the sample is in frozen helium at temperatures
near $T_{c}$, but, as seen in Fig. 5, no change in the pressure dependence
$T_{c}(P)$ is observed. \ This is not surprising since solid helium is the
softest solid known; in addition, the shear stresses are held to a minimum by
the carefully controlled manner \cite{schirber1} in which solid helium is
allowed to freeze around the sample.

A similar pressure derivative $dT_{c}/dP\simeq-1.07$ K/GPa to ours has very
recently been obtained by Lorenz \textit{et al.} \cite{lorenz2} in He-gas
studies to 0.8 GPa on a MgB$_{2}$ sample synthesized to stoichiometry with
superconducting midpoint at $T_{c}^{mid}(0)\simeq$ 39.2 K. \ These authors
also reexamined in a He-gas system the same sample studied earlier \cite{n14}
with $T_{c}^{mid}(0)\simeq$ 37.5 K and find $dT_{c}/dP\simeq-1.45$ K/GPa which
they report agrees within experimental error with their previous result
$dT_{c}/dP\simeq-1.6$ K/GPa to 1.8 GPa in a piston-cylinder device with
Fluorinert FC77 pressure medium. \ This appears to imply that the shear
stresses from frozen Fluorinert have little effect on the pressure dependence
of $T_{c}$ in the pressure range to 1.8 GPa.

Choi \textit{et al}. \cite{choi} have recently carried out resistivity studies
to 1.5 GPa pressure in daphne-kerosene pressure medium, obtaining
$dT_{c}/dP\simeq-1.36$ K/GPa. \ The results of all known $T_{c}(P)$
measurements on MgB$_{2}$ are summarized in the Table.

\subsubsection{$T_{c}(P)$ Measurements in the Diamond-Anvil System}

In Fig. 6 we show the dependence of $T_{c}$ on pressure to 20 GPa for
MgB$_{2}$ using a diamond-anvil-cell with dense helium pressure medium
\cite{n19}, thus extending the pressure range of the above He-gas studies
nearly thirtyfold. \ $T_{c}$ is seen to decrease nearly linearly with pressure
to 10 GPa, consistent with the rate -1.11 K/GPa (dashed line), but begins to
display a positive (upward) curvature at higher pressures. \ As will be
discussed below, this deviation originates from the increasing lattice
stiffness of MgB$_{2}$ at higher pressure. \ In these experiments the pressure
was always changed at RT, but measured at temperatures near $T_{c}$.%

\begin{center}
\includegraphics[
natheight=7.946800in,
natwidth=9.427300in,
height=3.5146in,
width=4.1649in
]%
{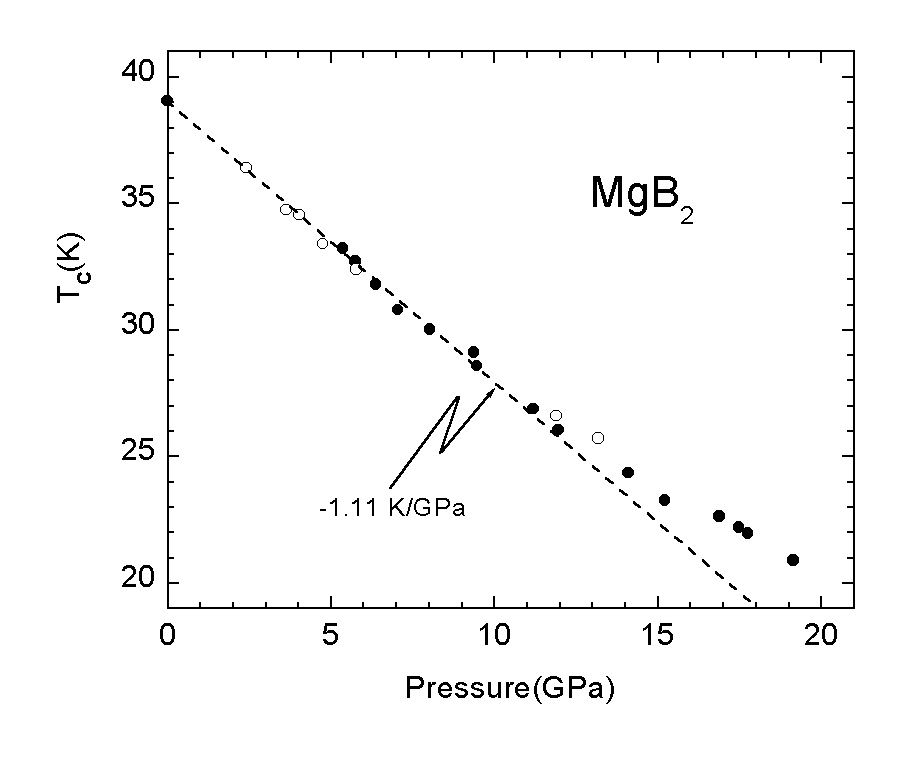}%
\end{center}

\noindent\textbf{Fig. 6.} \ Superconducting transition temperature midpoint
$T_{c}^{mid}$ versus pressure to 20 GPa from diamond-anvil-cell measurements
in Ref. \cite{n19}. \ Data with filled circles ($\bullet)$ taken for
monotonically increasing pressure, with open circles ($\circ$) for
monotonically decreasing pressure. \ The straight dashed line has slope -1.11
K/GPa.\vspace{0.6cm}

In Fig. 6 it is seen that the width of the superconducting transition
gradually increases from $\sim$ 0.3 K for P $\leq10$ GPa to 0.9 K at 19.2 GPa,
increasing somewhat further for the data with decreasing pressure. \ This
increase in width $\Delta T_{c}$ is seen to be usually accompanied by a slight
broadening of the ruby R1 fluorescence line; both broadening effects point to
a pressure gradient of approximately $\pm$ 0.3 GPa ($\pm$ 1.5\%) at the
highest pressures.\emph{\ }\ The magnitude of the shear stresses on the sample
would be expected to be larger in the diamond-anvil-cell than in the He-gas
experiment since the pressure range is much greater; above 12 GPa helium
freezes at RT so the diamonds must push on solid helium to increase the
pressure further. \ In addition, in the diamond-anvil-cell it is not possible
to cool slowly through the melting curve of helium with a well-defined
temperature gradient. \ However, the data in Fig. 6 give no clear indication
for shear stress effects on $T_{c}$ at any pressure.

Very recently Tissen \textit{et al.} \cite{tissen3} have carried out ac
susceptibility measurements in a diamond-anvil-cell to 28 GPa on a MgB$_{2}$
sample with $T_{c}^{mid}(0)\simeq$ 37.3 K at ambient pressure. \ They find an
initial slope $dT_{c}/dP\simeq-2$ K/GPa, $T_{c}$ decreasing to 11 K at 20 GPa
and 6 K at 28 GPa, a 50\% greater decrease than observed by either us (see
Fig. 6) or Monteverde \textit{et al.} \cite{monte}. \ They also report that
the pressure dependence $T_{c}(P)$ shows a bump near 9 GPa which they
speculate may arise from an electronic Lifshitz transition. \ We suggest that
shear stress effects may also play a role in their measurements. \ At 20 GPa
the width in their superconducting transition has increased by $\sim$ 3 K
which would correspond to a pressure gradient of $\sim$ 3.5 GPa, an order of
magnitude higher than in our helium-loaded diamond-anvil-cell measurements.

The degree to which shear stresses affect the data of Monteverde \textit{et
al. }\cite{monte} is unknown. \textit{\ }However, since shear stresses are
potentially much larger in solid pressure media such as steatite than in
frozen fluids such as helium or Fluorinert, it would seem likely that they are
responsible for at least part of the widely differing $T_{c}(P)$ dependences
to 25 GPa observed in three of their four experiments.As discussed in the
Introduction, the initial pressure derivative $dT_{c}/dP$ in the present
experiment (-1.11 K/GPa) differs significantly from those (-0.3 to -2.0 K/GPa)
obtained by other groups using pressure media which are either solid at RT or
readily freeze upon cooling \cite{monte,n14,n14',choi,tissen3}. \ It is not
yet clear whether these widely varying results reflect differences in the
make-up of the samples or differing degrees of shear stress exerted on the
samples by the various frozen or solid pressure media. \ An inspection of the
data in the Table suggests a possible correlation voiced by Tissen \textit{et
al.} \cite{tissen3} that larger values of $\left|  dT_{c}/dP\right|  $ are
associated with lower ambient-pressure values of $T_{c}$. \ However, it is
difficult to accurately compare $T_{c}$ values determined in ac susceptibility
and electrical resistivity measurements, the latter usually lying higher; in
addition, in the ac susceptibility the value of $T_{c}$ may depend somewhat on
the ac field strength. \ Further experimentation under carefully controlled
conditions is clearly necessary to investigate this possible correlation.\newpage

\noindent\textbf{Table. \ }Summary of available high-pressure $T_{c}(P)$ data
on MgB$_{2}$. \ $T_{c}$ values are at ambient pressure from superconducting
midpoint in ac susceptibility $\chi_{ac}$ and electrical resistivity $\rho$
measurements. \ $dT_{c}/dP$ is initial pressure derivative. \ $P^{\max}$(GPa)
is the maximum pressure reached in experiment .\vspace{0.2cm}

\noindent%
\begin{tabular}
[c]{|l|l|l|l|l|l|}\hline
$T_{c}$(K) & $\frac{dT_{c}}{dP}$(K/GPa) & $P^{\max}$(GPa) &
\textbf{measurement} & $%
\begin{array}
[c]{c}%
\text{\textbf{pressure}}\\
\text{\textbf{medium}}%
\end{array}
$ & \textbf{reference}\\\hline\hline
39.1 & -1.1 & 19.2 & $\chi_{ac}$, $^{11}$B isotope & helium & Fig. 6
\cite{n19}\\\hline
39.1 & -1.11(2) & 0.66 & $\chi_{ac}$, $^{11}$B isotope & helium & Fig. 5
\cite{n18}\\\hline
39.1 & -1.09(4) & 0.63 & $\chi_{ac}$, $^{11}$B isotope & helium &
\cite{hamlin}\\\hline
39.2 & -1.11(3) & 0.61 & $\chi_{ac}$, $^{11}$B isotope & helium &
\cite{hamlin}\\\hline
40.5 & -1.12(3) & 0.64 & $\chi_{ac}$, $^{10}$B isotope & helium &
\cite{hamlin}\\\hline
39.2 & -1.07 & 0.84 & $\chi_{ac}$ & helium & \cite{lorenz2}\\\hline
37.4 & -1.45 & 0.84 & $\chi_{ac}$ & helium & \cite{lorenz2}\\\hline
37.4 & -1.6 & 1.84 & $\chi_{ac}$ & Fluorinert FC77 & \cite{n14}\\\hline
37.3 & -2 & 27.8 & $\chi_{ac}$ & 4:1 meth.-ethanol & \cite{tissen3}\\\hline
38.2 & -1.36 & 1.46 & $\rho$ & daphne-kerosene & \cite{choi}\\\hline
37.5 & -1.9 & 1.35 & $\rho$ & Fluorinert FC70 & \cite{n14'}\\\hline
$\sim$ 35 & -0.35 to -0.8 & 25 & $\rho$ & steatite, RT solid & \cite{monte}%
\\\hline
\end{tabular}

\section{DISCUSSION}

The present studies of both the superconducting and structural properties of
MgB$_{2}$ under hydrostatic pressure were carried out on the same high quality
MgB$_{2}$ sample used in the He-gas measurements to 0.7 GPa. \ These combined
studies thus allow an accurate determination of the change in $T_{c} $ with
unit cell volume $V$ for comparison with theory. \ The change in $T_{c}$ with
$V$ is given by
\begin{equation}
\frac{d\ln T_{c}}{d\ln V}=\frac{B}{T_{c}}\left(  \frac{dT_{c}}{dP}\right)
=+4.16(8),
\end{equation}
using the above values $dT_{c}/dP\simeq-1.11(2)$ K/GPa, $B=147.2(7)$ GPa, and
$T_{c}=39.25$ K. \ This value of $d\ln T_{c}/d\ln V$ is somewhat smaller than
that (+6.6) obtained by Neaton and Perali \cite{neaton} in an estimate based
on density functional theory.

We will now discuss the implications of this result for the nature of the
superconducting state in MgB$_{2}.$ \ First consider the McMillan equation
\cite{r17}
\begin{equation}
T_{c}\simeq\frac{\left\langle \omega\right\rangle }{1.20}\exp\left\{
\frac{-1.04(1+\lambda)}{\lambda-\mu^{\ast}(1+0.62\lambda)}\right\}  ,
\end{equation}
valid for strong coupling ($\lambda\lesssim1.5),$ which connects the value of
$T_{c}$ with the electron-phonon coupling parameter $\lambda,$ an average
phonon frequency $\left\langle \omega\right\rangle ,$ and the Coulomb
repulsion $\mu^{\ast},$ which we assume to be pressure independent
\cite{chen2}. \ The coupling parameter is defined by $\lambda=N(E_{f}%
)\left\langle I^{2}\right\rangle /[M\left\langle \omega^{2}\right\rangle ],$
where $N(E_{f})$ is the electronic density of states at the Fermi energy,
$\left\langle I^{2}\right\rangle $ the average squared electronic matrix
element, $M$ the molecular mass, and $\left\langle \omega^{2}\right\rangle $
the average squared phonon frequency. Taking the logarithmic volume derivative
of $T_{c}$ in Eq. (3), we obtain the simple relation
\begin{equation}
\frac{d\ln T_{c}}{d\ln V}=-\gamma+\Delta\left\{  \frac{d\ln\eta}{d\ln
V}+2\gamma\right\}  ,
\end{equation}
where $\gamma\equiv-d\ln\left\langle \omega\right\rangle /d\ln V$ is the
Gr\"{u}neisen parameter, $\eta\equiv N(E_{f})\left\langle I^{2}\right\rangle $
is the Hopfield parameter \cite{r17''}, and
\begin{equation}
\Delta\equiv\frac{1.04\lambda\lbrack1+0.38\mu^{\ast}]}{\left[  \lambda
-\mu^{\ast}(1+0.62\lambda)\right]  ^{2}}.
\end{equation}
Eq. (4) has a simple interpretation. \ The first term on the right, which
comes from the prefactor to the exponent in the above McMillan expression for
$T_{c}$, is usually small relative to the second term, as will be shown below.
\ The sign of the logarithmic derivative $d\ln T_{c}/d\ln V$, therefore, is
determined by the relative magnitude of the two terms in the curly brackets.

The second ``electronic'' term in Eq. (4) involves the logarithmic volume
derivative of the Hopfield parameter $\eta\equiv N(E_{f})\left\langle
I^{2}\right\rangle $, an ``atomic'' property which can be calculated directly
in band-structure theory \cite{r17}. \ In his landmark paper \cite{r17},
McMillan demonstrated that whereas $N(E_{f})$ and $\left\langle I^{2}%
\right\rangle $ individually may fluctuate appreciably as one element is
substituted for another across a transition-metal alloy series and the
d-electron count varies, their product $\eta\equiv N(E_{f})\left\langle
I^{2}\right\rangle $ changes only gradually, i.e. $\eta$ is a well behaved
``atomic'' property. \ One would thus anticipate that $\eta$ changes in a
relatively well defined manner under pressure, reflecting the character of the
electrons near the Fermi energy. \ An examination of the body of high-pressure
data on simple s,p-metal superconductors, in fact, reveals that $\eta$
normally increases under pressure at a rate given by $d\ln\eta/d\ln
V\approx-1$ \cite{r17'}. \ For transition-metal (d-electron) superconductors,
on the other hand, Hopfield has pointed out that $d\ln\eta/d\ln V\approx-3$ to
-4 \cite{r17''}.

The second ``lattice'' term in the curly brackets in Eq. (4) is positive,
typically $2\gamma\approx3-5.$ \ Since in simple metal superconductors, like
Al, In, Sn, and Pb, this positive ``lattice'' term dominates over the
electronic term $d\ln\eta/d\ln V\approx-1$, and\ $\Delta$ is always positive,
the sign of $d\ln T_{c}/d\ln V$ is the same as that in the curly brackets,
namely positive; this accounts for the universal decrease of $T_{c} $ with
pressure due to lattice stiffening in simple metals. \ In selected transition
metals the electronic term may become larger than the lattice term, in which
case $d\ln T_{c}/d\ln V$ is negative and $T_{c}$ would be expected to
\textit{increase} with pressure, as observed, for example, in experiments on V
\cite{smith1} and La \cite{smith2}.

Let us now apply Eq. (4) in more detail to a canonical BCS simple-metal
superconductor. $\ $In Sn, for example, $T_{c}$ decreases under pressure at
the rate $dT_{c}/dP\simeq$ -0.482 K/GPa which leads to $d\ln T_{c}/d\ln
V\simeq+7.2$ \cite{r20}. \ We note that this value of $d\ln T_{c}/d\ln V$ is
almost twice as large as that for MgB$_{2}$ (see Eq. (2)); this is exactly
what is expected from Eq. (4) since $\Delta$ increases for \textit{decreasing}
values of $T_{c}.$ \ Inserting for Sn $T_{c}(0)\simeq$ 3.73 K, $\left\langle
\omega\right\rangle \simeq110$ K \cite{r21}, and $\mu^{\ast}=0.1$ into the
above McMillan equation, we obtain $\lambda\simeq0.69$ from which follows that
$\Delta\simeq2.47.$ \ Inserting the above values into Eq. (4) and setting
$d\ln\eta/d\ln V\approx-1$ for simple metals, we can solve Eq. (4) for the
Gr\"{u}neisen parameter to obtain $\gamma\simeq+2.46,$ in reasonable agreement
with experiment for Sn ($\gamma\approx+2.1) $ \cite{r20}. \ Similar results
are obtained for other conventional simple metal BCS superconductors.

We now repeat the same calculation with the McMillan equation for MgB$_{2}$
using the logarithmically averaged phonon energy from inelastic neutron
studies \cite{n9} $\left\langle \omega\right\rangle =670$ K, $T_{c}%
(0)\simeq39.25$ K, and $\mu^{\ast}=0.1,$ yielding $\lambda\simeq0.90$ and
$\Delta\simeq1.75$ from Eqs. (3) and (5), respectively. \ Our estimate of
$\lambda\simeq0.90$ agrees well with those of other authors \cite{kong,an}.
\ Since the pairing electrons in MgB$_{2}$ are believed to be s,p in character
\cite{kortus,medvedeva,kong,neaton}, we set $d\ln\eta/d\ln V\approx-1,$ a
value close to $d\ln\eta/d\ln V=Bd\ln\eta/dP\approx-0.81,$ where $B=147.2$ GPa
from Ref. \cite{r10} and $d\ln\eta/dP\approx+0.55$ \%/GPa from
first-principles electronic structure calculations by Medvedera \textit{et
al.} \cite{note10}. \ Inserting the values of $d\ln T_{c}/d\ln V=+4.16$,
$\Delta=1.75$, and $d\ln\eta/d\ln V=-1$ into Eq. (4), we find $\gamma
\simeq2.36,$ in reasonable agreement with the value $\gamma\approx2.9$ from
Raman spectroscopy studies \cite{goncharov} or $\gamma\approx2.3$ from
\textit{ab initio }electronic structure calculations on MgB$_{2}$ \cite{r18}.

In spite of the significant compression anisotropy, electronic structure
calculations based on the high-pressure structural data show that the
electronic structure does not change much at high pressure \cite{note10}; the
calculations show that the electric field gradient in MgB$_{2}$ is essentially
independent of pressure up to 10 GPa. \ As the electric field gradient is a
very sensitive characteristic of the electronic charge distribution, one may
conclude that no large changes in the partial charges of the B $2p$ states and
boron electronic structure take place under pressure. \ Further results from
theory support this conclusion. \ Medvedera \textit{et al.} \cite{note10} find
the Hopfield parameter for MgB$_{2}$ to only depend weakly on pressure
$d\ln\eta/dP\approx+0.55$ \%/GPa. \ The change in the electronic density of
states $d\ln N(E_{f})/dP$ is also estimated to be very small: \ Loa and
Syassen \cite{loa} (-0.31 \%/GPa), Medvedera \textit{et al.} \cite{note10}
(-0.51 \%/GPa), and Vogt et al. \cite{r9} (-0.38 \%/GPa). \ Assuming $B=147.2$
GPa, one thus obtains $d\ln N(E_{f})/d\ln V\simeq$ +0.46, +0.75, and +0.56,
respectively. \ These values are near that (+0.67) expected for a 3D free
electron gas. \ Since $d\ln\eta/dP=d\ln N(E_{f})/dP+d\ln\left\langle
I^{2}\right\rangle /dP,$ these results imply that the average squared
electronic matrix element $\left\langle I^{2}\right\rangle $ in MgB$_{2}$
increases under pressure at the approximate rate of only +1 \%/GPa. \ The sign
and magnitude of the changes in $N(E_{f})$ and $\left\langle I^{2}%
\right\rangle $ under hydrostatic pressure for MgB$_{2}$ are comparable to
those found for simple s,p-metal superconductors. \ Larger changes are
anticipated if uniaxial pressure is applied \cite{uniaxial}. \ The main reason
for the observed decrease of $T_{c}$ with pressure is not an electronic
effect, but a strong pressure enhancement of the phonon frequencies, an effect
which has been directly observed in Raman measurements \cite{goncharov}.

Taken as a whole, the above results thus give considerable evidence that the
superconducting state of MgB$_{2}$ is strongly related to that in simple
s,p-metal superconductors like Al, Sn, In, and Pb which exhibit BCS
phonon-mediated superconductivity. \ This is not to say that superconductivity
in MgB$_{2}$ is identical to that in the simple metals. \ Extensive specific
heat \cite{n8'} and high-resolution photoemission studies \cite{tsuda} on
MgB$_{2}$ give evidence for a multicomponent superconducting gap.

The above analysis is based on the results of the present high-pressure
studies using the He-gas technique to 0.7 GPa . \ We now consider the
diamond-anvil-cell data to 20 GPa in Fig. 6. \ For comparison to theory it is
advantageous to use the Murnaghan equation-of-state to convert pressure to
relative volume $V/V_{0}$
\begin{equation}
\frac{V(P)}{V_{0}}=\left[  1+\frac{B^{\prime}P}{B}\right]  ^{-1/B^{\prime}},
\end{equation}
where we use the value $B=147.2$ GPa from Ref. \cite{r10} and the canonical
value $B^{\prime}\equiv dB/dP=4$ supported by a recent calculation \cite{loa}.
\ In Fig. 7 we replot the data from Fig. 6 as $T_{c}$ versus relative volume
$V/V_{0}.$ \ The maximum pressure applied in the present experiment (19.2 GPa)
results in a volume decrease of $\sim$ 10\%. \ Much of the nonlinearity in the
$T_{c}$ versus pressure plot in Fig. 6 appears to disappear when $T_{c}$ is
plotted versus $V/V_{0}$.

We now compare the $T_{c}$ versus $V/V_{0}$ dependence in Fig. 7 to the result
from the He-gas data which yields the initial volume dependence $d\ln
T_{c}/d\ln V\simeq$ +4.16 given in Eq. (2). \ If we assume this relation holds
at all pressures, then we can integrate it to obtain
\begin{equation}
\frac{T_{c}}{(39.25\text{ K})}=\left(  \frac{V}{V_{0}}\right)  ^{+4.16},
\end{equation}
which is plotted as the upper solid line in Fig. 7. \ This volume dependence
must be accurate for small pressures where $V/V_{0}\simeq1,$ corresponding to
the pressure dependence $dT_{c}/dP=-1.11$ K/GPa from the He-gas data, but
rises well above the experimental data at higher pressures.

Another way to extrapolate the He-gas data to higher pressures is to assume
that $T_{c}$ varies linearly with volume change $\Delta V,$ yielding from Eq.
(7)
\begin{equation}
\frac{T_{c}}{39.25\text{ K}}=\left(  \frac{V_{0}+\Delta V}{V_{0}}\right)
^{+4.16}\simeq\left(  1+4.16\frac{\Delta V}{V_{0}}\right)  =\left(
-3.16+4.16\frac{V}{V_{0}}\right)  ,
\end{equation}
which is plotted as the straight dashed line in Fig. 7. \ As they must, the
upper solid and dashed lines agree exactly near $V/V_{0}=1.$ \ The dashed line
is seen to lie above the experimental data points at higher pressures and to
extrapolate to $T_{c}=$ 0 K for $V/V_{0}=0.76$ which corresponds to an applied
pressure of $\sim$ 75 GPa. \ A least-squares straight line fit through all
data in Fig. 7 leads to the estimate that $T_{c}=$ 0 K for $P\approx$ 60 GPa.%

\begin{center}
\includegraphics[
natheight=8.060100in,
natwidth=9.693700in,
height=3.5146in,
width=4.2211in
]%
{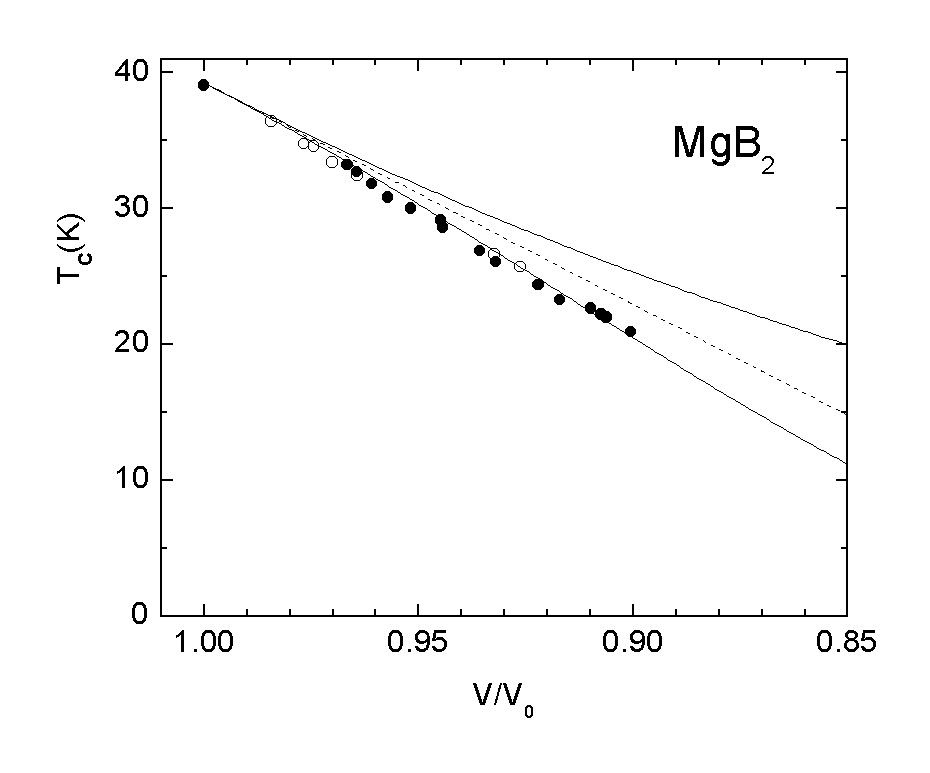}%
\end{center}

\noindent\textbf{Fig. 7.} \ \ $T_{c}$ data from figure 6 plotted versus
relative volume $V/V_{0}$ (from Ref. \cite{n19}). \ See text for explanation
of solid and dashed lines.\vspace{0.6cm}

It is not surprising that $T_{c}$ is not a linear function of $V/V_{0}$ to
very high pressures. \ In the McMillan formula in Eq. (3) $T_{c}$ depends
exponentially on the solid state parameters and it is the relatively small
changes in these parameters which lead to the large change in $T_{c}$ under
pressure see in Figs. 6 and 7. \ As pointed out by Chen \textit{et al.}
\cite{chen2}, a more appropriate method to estimate the dependence of $T_{c}$
on relative volume $V/V_{0}$ would thus be to integrate the volume derivatives
of these parameters $\gamma\equiv-d\ln\left\langle \omega\right\rangle /d\ln
V=$ +2.36, $d\ln\lambda/d\ln V=d\ln\eta/d\ln V-d\ln\left\langle \omega
^{2}\right\rangle /d\ln V=-1-2(-2.36)=+3.72$ to obtain $\left\langle
\omega\right\rangle =($670 K$)(V/V_{0})^{-2.36}$ and $\lambda=0.90(V/V_{0}%
)^{3.72}.$ \ Inserting these two volume dependences in the McMillan equation,
and assuming $\mu^{\ast}=0.1$ is independent of pressure \cite{chen2}, we
obtain the dependence of $T_{c}$ on relative volume shown as the lower solid
line in Fig. 7. \ The agreement with the experimental data is quite
impressive. \ Note that according to this estimate approximately 50 GPa
pressure would be required to drive $T_{c}$ to below 4 K. \ A similar
calculation was very recently carried out by Chen \textit{et al.} \cite{chen2}
over a much wider pressure range; this paper also contains a detailed
discussion of the pressures dependences of $\eta,$ $\lambda,$ and $\mu^{\ast
}.$ \ The good agreement between the experimental data to 20 GPa and the
predictions of the McMillan formula using the volume dependences determined
from the He-gas high-pressure data to 0.7 GPa lends additional evidence that
superconductivity in MgB$_{2}$ originates from standard BCS phonon-mediated
electron pairing.

In conventional metals, electron-phonon scattering makes the dominant
contribution to the temperature-dependent electrical resistivity $\rho(T).$
\ At sufficiently high temperatures, Bloch-Gr\"{u}neisen \cite{bloch} theory
gives a linear dependence on temperature $\rho_{RT}=bT,$ where $b\propto
r_{s}^{-2}\Theta_{D}^{-2},$ $r_{s}$ is the radius of the Wigner-Seitz sphere,
and $\Theta_{D}$ is the Debye temperature. \ Near RT Choi \textit{et al.}
\cite{choi} find the electrical resistivity of MgB$_{2}$ to increase linearly
with temperature. \ Under pressure these authors find that the RT electrical
resistivity decreases under pressure at the rate $d\ln\rho_{RT}/dP\simeq-3$
\%/GPa. \ Using the bulk modulus $B=147.2$ GPa, this yields $d\ln\rho
_{RT}/d\ln V\simeq$ +4.42. \ Taking the logarithmic volume derivative of the
above Bloch-Gr\"{u}neisen expression and using the free-electron expression
for $r_{s}$ and setting $\gamma=2.36$ from above, we obtain \ $d\ln\rho
_{RT}/d\ln V=2\gamma-2/3=+4.05,$ in surprisingly good agreement with the
measured value. \ It is significant that the same value of the Gr\"{u}neisen
parameter yields the pressure dependence of the electron-phonon interaction
which accounts for both $T_{c}(P)$ and $\rho_{RT}(P).$

At first glance the present results appear to be inconsistent with the hole
superconductivity model of Hirsch and Marsiglio \cite{r7,r7'} which predicts
that $T_{c}$ should increase with pressure if there is no change in the doping
level of holes. \ Indeed, the pressure-induced change in the concentration of
hole-carriers in the boron $\sigma$-band is estimated to be extemely small
\cite{loa,note10}. \ Further experiments, such as high-pressure Hall effect
measurements, are necessary to determine what, if any, change in the carrier
concentration occurs. \ The success of the above analysis of the dependence of
$T_{c}$ on pressure gives further evidence that MgB$_{2}$ is an extraordinary
superconductor which makes the most out of its conventional BCS
electron-phonon pairing interaction.\vspace{0.4cm}

\noindent\textbf{Acknowledgments}

The authors are grateful to S. Short for assistance with the preparation of
figures for the paper and to X. J. Chen for providing a preprint of his recent
paper. \ The authors would like to thank S. Klotz for providing the ruby
spheres. \ Work at Washington University supported by NSF grant DMR-0101809
and that at the Argonne National Laboratory by the U.S. Department of Energy,
Office of Science, contract No. W-31-109-ENG-38.

\begin{thebibliography}{9}
\bibitem {klotz}The ruby spheres, which were annealed at 1500$^{\circ}$C for
one week followed by a cooldown to RT over a three-week period, were prepared
by J. C. Chervin, Physique des Milieux Condens\'{e}s, Paris, and contained
3600 ppm Cr$^{3+}.$ \
\end{thebibliography}
\end{document}